\documentclass[preprint,times]{aastex}
\usepackage{natbib}

\usepackage{graphicx}
\usepackage{epsf}

\begin{document}

\title{Identification of Magnetite in B-type Asteroids}
\author{Bin Yang$^{1}$ and David Jewitt$^{2}$
\affil{$1$ Institute for Astronomy, University of Hawaii, Honolulu, HI
96822 \\ 
$2$ Department of Earth and Space Sciences, Institute for Geophysics and Planetary Physics and Department of Physics and Astronomy,  UCLA, Los Angeles, CA 90095} 
\email{yangbin@ifa.hawaii.edu, jewitt@ucla.edu}}

\begin{abstract}
Spectrally blue (B-type) asteroids are rare, with the second discovered asteroid Pallas being the largest and most famous example. We conducted a focused, infrared spectroscopic survey of B-type asteroids to search for water-related features in these objects. Our results show that the negative optical spectral slope of some B-type asteroids is due to the presence of a broad absorption band centered near 1.0 $\mu$m. The 1-$\mu$m band can be matched in position and shape using magnetite (Fe$_{3}$O$_{4}$), which is an important indicator of past aqueous alteration in the parent body. Furthermore, our observations of B-type asteroid (335) Roberta in the 3-$\mu$m region reveal an absorption feature centered at 2.9 $\mu$m, which is consistent with the absorption due to phyllosilicates (another hydration product) observed in CI chondrites. The new observations suggest that at least some B-type asteroids are likely to have incorporated significant amounts of water ice and to have experienced intensive aqueous alteration. 
\end{abstract}

\section{Introduction}
Many meteoroid streams are associated with the orbits of active comets (Jenniskens 2008). For instance, the Eta Aquarids and the Orionids are associated with comet Halley \citep{whipple:1951,mckinley:1961}, while the Leonid shower and the Perseid shower are associated with comet 55P/Tempel-Tuttle \citep{Mason:1995} and comet 109P/Swift-Tuttle \citep{Jenniskens:1998}, respectively. However, the Geminid shower is instead associated with (3200) Phaethon  \citep{whipple:1983}, whose orbit is that of an asteroid and which has failed to show any evidence of on-going mass loss (Chamberlin et al.~1996, Hsieh and Jewitt 2005). It is worth mentioning here that the orbit of Phaethon is unusual even among  near earth asteroids (NEAs) in having a very short period (P $\sim$ 1.43 yr) along with a small perihelion distance of $q$ = 0.141 AU \citep{green:1983}. What makes it more interesting is that Phaethon shows a conspicuously bluish (negatively sloped)  0.4 to 0.9 $\mu$m spectrum \citep{tholen:1984, Luu:1990, chamberlin:1996, lazzarin:1996, licandro:2007}. Blue asteroids are relatively rare, with only $\sim$ 4\% being bluer than the solar color \citep{tholen:1984, bus:2002, binzel:2002, dandy:2003}. These rare blue objects are classified as B-type in the Bus taxonomic system \citep{bus:1999}.  

Previous studies have recognized that Phaethon is not the only blue object that is found to be associated with past cometary activity \citep{meng:2004, licandro:2007, licandro:2009}. For example, \cite{ohtsuka:2006} identified an Apollo-type NEA 2005 UD that shows a orbit similar to that of 3200 Phaethon. Broadband photometry of 2005 UD revealed that it has blue optical colors that resemble those of Phaethon, consistent with the hypothesis that these two bodies have a common origin \citep{jewitt:2006, kinoshita:2007}. Moreover, another Apollo asteroid 1999 YC was found to be dynamically associated with 3200 Phaethon and 2005 UD \citep{ohtsuka:2008}. These authors further suggested the existence of a ``Phaethon-Geminid stream Complex (PGC)", which consists of a group of fragments that share similar dynamical properties \citep{kasuga:2008}. Beside the Phaethon-Geminid Stream, NEA 2001 YB$_5$ \citep{meng:2004} and 4015 Wilson-Harrington (also known as comet 107P; \cite{cunningham:1950}, \cite{fernandez:1997}) are reported to be the possible parent bodies of weak meteor showers.  Spectra of both 2001 YB$_5$ \citep{yang:2003} and 4015 Wilson-Harrington \citep{tholen:1984} are blue/neutral in the optical and are classified as Bus B-types.  

The over-abundance of B-type reflection spectra in asteroids inferred to have been active in the past is unlikely to be due to chance. On the other hand, some B-types (like Pallas itself) show no hint of past activity, and only $\sim$15\% of (mostly Jupiter family) cometary nuclei show blue optical colors (Jewitt 2002), so the meaning of this association is unclear. The present paper was motivated by the possibility that some B-types might have a common origin, perhaps related to water-rich compositions. Accordingly, we undertook spectral observations of B-type asteroids to search for evidence of water ice or other volatiles. 

\section{Near Infrared Spectroscopy}
\subsection{Observation and Data Reduction}
Near infrared (NIR) observations were taken in May 2008, using the NASA Infrared Telescope Facility (IRTF) 3-m telescope atop Mauna Kea, Hawaii. A medium-resolution 0.8-5.5 $\mu$m spectrograph (SpeX) was used, equipped with a Raytheon 1024 x 1024 InSb array having a spatial scale of 0.$\!\!^{\prime\prime}$15 pixel$^{-1}$ \citep{rayner:2003}. The low-resolution prism mode gave an overall wavelength range from 0.8 $\mu$m to 2.5 $\mu$m for all of our NIR observations. We used a 0.8$^{\prime\prime}$ x 15$^{\prime\prime}$ slit that provided an average spectral resolving power of $\sim$ 130.  To remove sky background, an ABBA dither pattern was used with an interval of 7$^{\prime\prime}$ along the slit. Seeing was below 1$^{\prime\prime}$ for all the nights in this study.  During our observations, the slit was always oriented along the parallactic angle to minimize effects from differential atmospheric refraction. 

In order to correct for strong telluric absorption features from oxygen, water vapor and other atmospheric species, we observed G2V stars close to the scientific target both in time and in sky position. The G2V stars generally have spectra similar to the Sun and they also serve as solar analogs for computing relative reflectance spectra of scientific targets. The SpeX data were reduced using the SpeXtool reduction pipeline \citep{cushing:2004}. Other details of data reduction were described in a previous paper \citep{yang:2009}. 

A journal of observations is provided  in Table \ref{obs}. 

\subsection{Results and Analysis}
\begin{figure}[h]
\vspace{0.5 cm}
\hspace{-1 cm}\includegraphics[width=7in,angle=0]{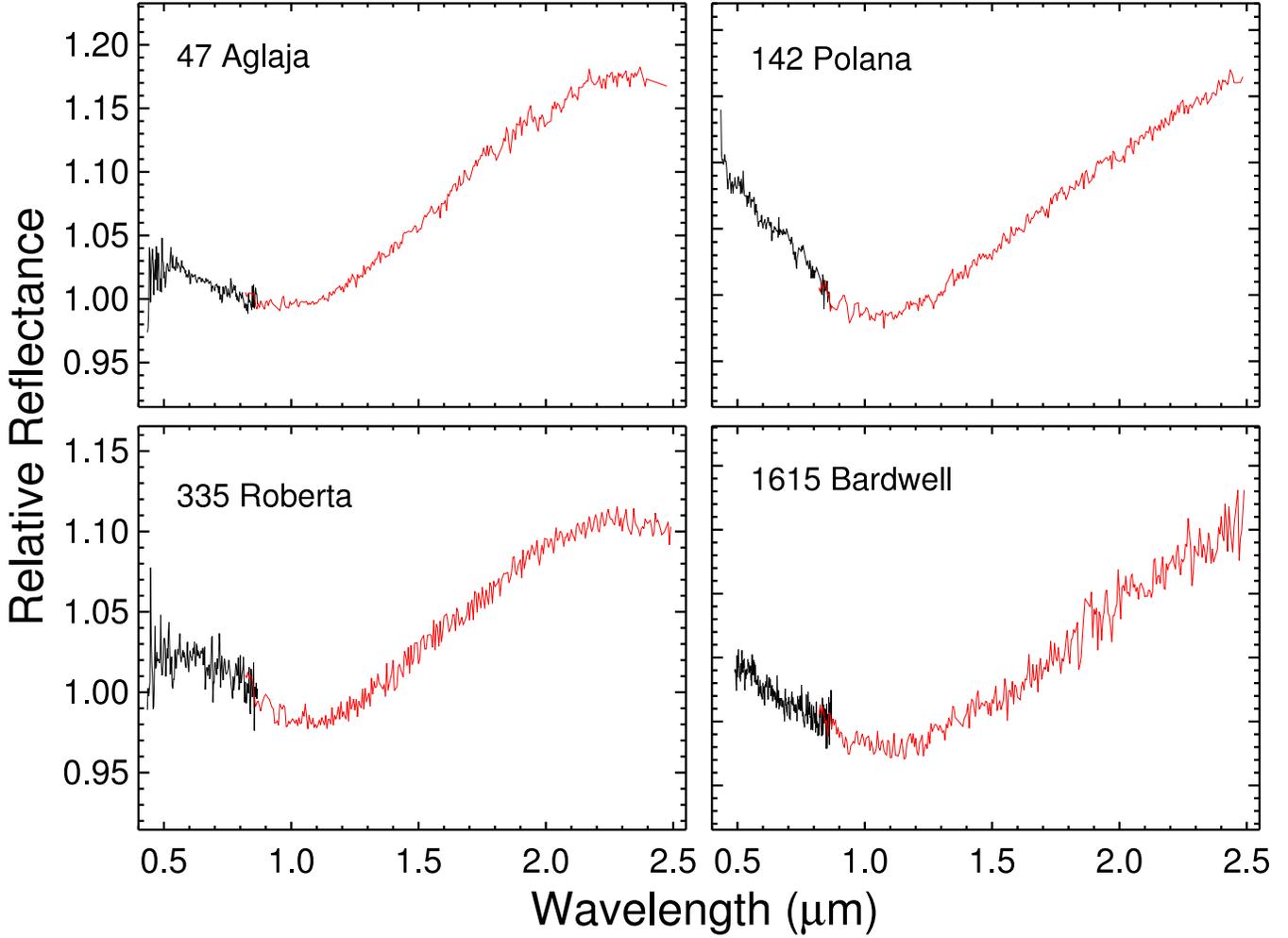}
\caption{Visible and NIR spectra of 4 B-type asteroids that show a broad absorption band near 1.0 $\mu$m. The visible data are shown in black and the NIR spectra obtained in this study are shown in red. The visible observations of  asteroid (1615) Bardwell) is taken from the S3oS2 survey \citep{lazzaro:2004} and the visible data of the other 3 objects are from the SMASS II survey \citep{bus:2002}. The visible and the NIR spectra are normalized at 0.85 $\mu$m.}
\label{f1}
\end{figure}

In order to study B-type asteroids over a wide wavelength range, we combined our NIR (0.8 - 2.5 $\mu$m) spectra with available visible (0.44 - 0.85 $\mu$m) observations for all the objects in our sample. The visible spectra were taken from two spectroscopic surveys, namely the SMASS II survey \citep{bus:2002} and the S3OS2 survey \citep{lazzaro:2004}. We normalized and merged the NIR data and the visible data at 0.85 $\mu$m and then trimmed off the excess wavelength coverage. When combining the NIR and the visible data, we found that many B-type asteroids in our sample show a broad absorption band in their reflectance spectra with the minimum reflectivity near 1-$\mu$m. The band center is far from the region of overlap between the visible and near infrared spectra and thus cannot be an artifact of combining different data. The merged spectra of four objects that exhibit typical 1-$\mu$m absorption bands are shown in Figure \ref{f1}, in which the black solid lines are the visible data and the red solid lines are the NIR observations obtained in this study. Figure \ref{f1} clearly shows that the blueness of these B-type asteroids, in the optical, is the result  of blue-wing absorption from the 1-$\mu$m band. 

Physical and dynamical properties of the four objects presented in Figure \ref{f1} are listed in Table \ref{orb_phy}. Although the four B-type asteroids show similar spectral profiles both in the optical and the NIR, they have noticeably different orbits as well as different albedos and sizes. Specifically, (142) Polana is a member of the Polana dynamical family \citep{Cellino:2001}, (1615) Bardwell is a member of the Themis dynamical family \citep{Zappala:1995} which hosts the two well-known main belt comets, 133P/Elst-Pizarro and 176P/Linear \citep{hsieh:2006}. The other two, (47) Aglaja and (335) Roberta, do not belong to any identified dynamical family. To further characterize compositional properties of the B-type asteroids and to investigate the nature of the 1-$\mu$m absorption band, spectral models were applied. Since the four objects are spectrally very similar, we focus our discussion of the modeling results on (335) Roberta with the understanding that eventually similar conclusions apply to the other three asteroids. 

\subsubsection{Linear Spectral Mixing Model}
\begin{figure}[h]
\begin{center}
\vspace{-1.0 cm}
\includegraphics[width=4.5in,angle=0]{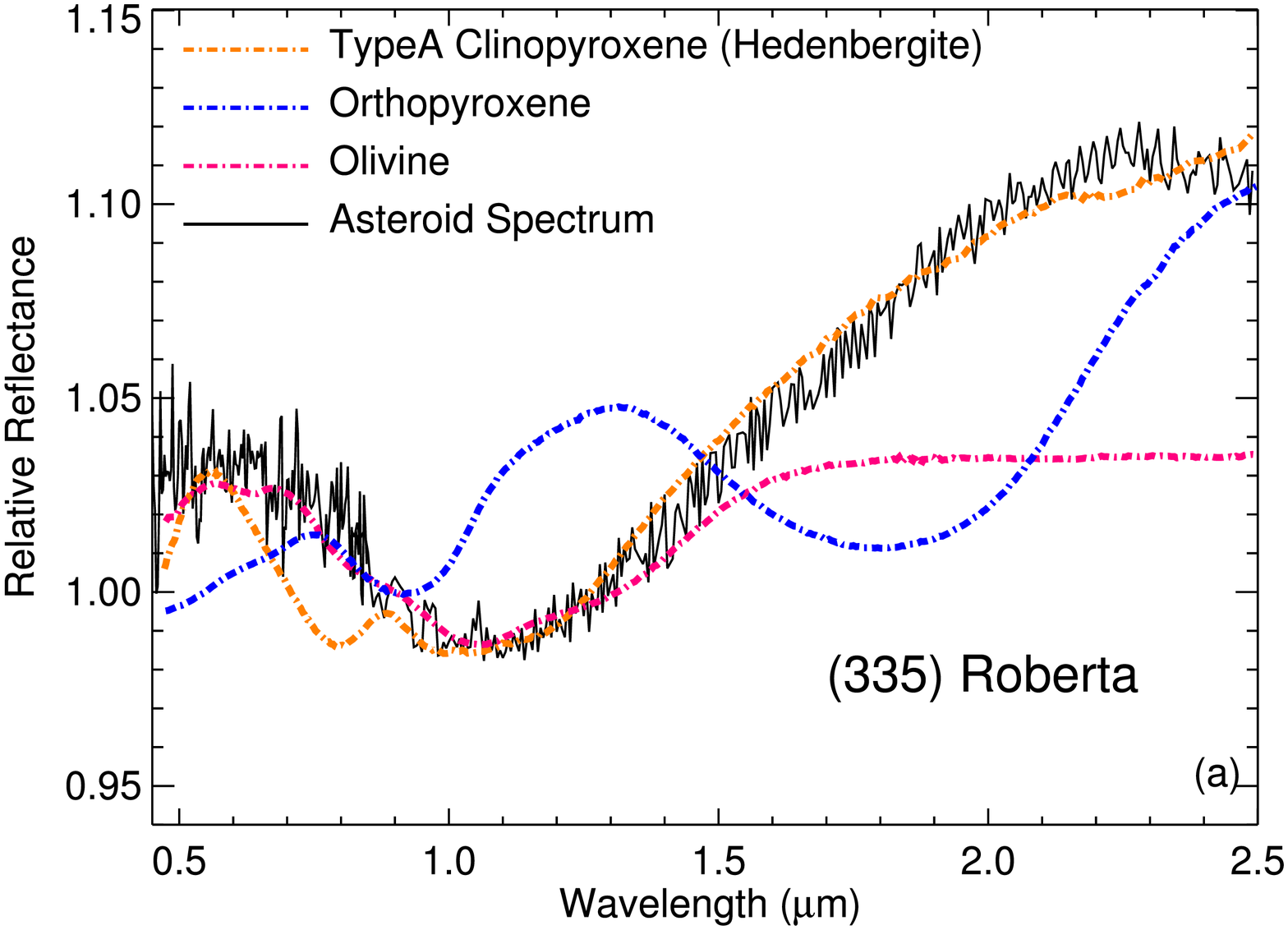}
\includegraphics[width=4.5in,angle=0]{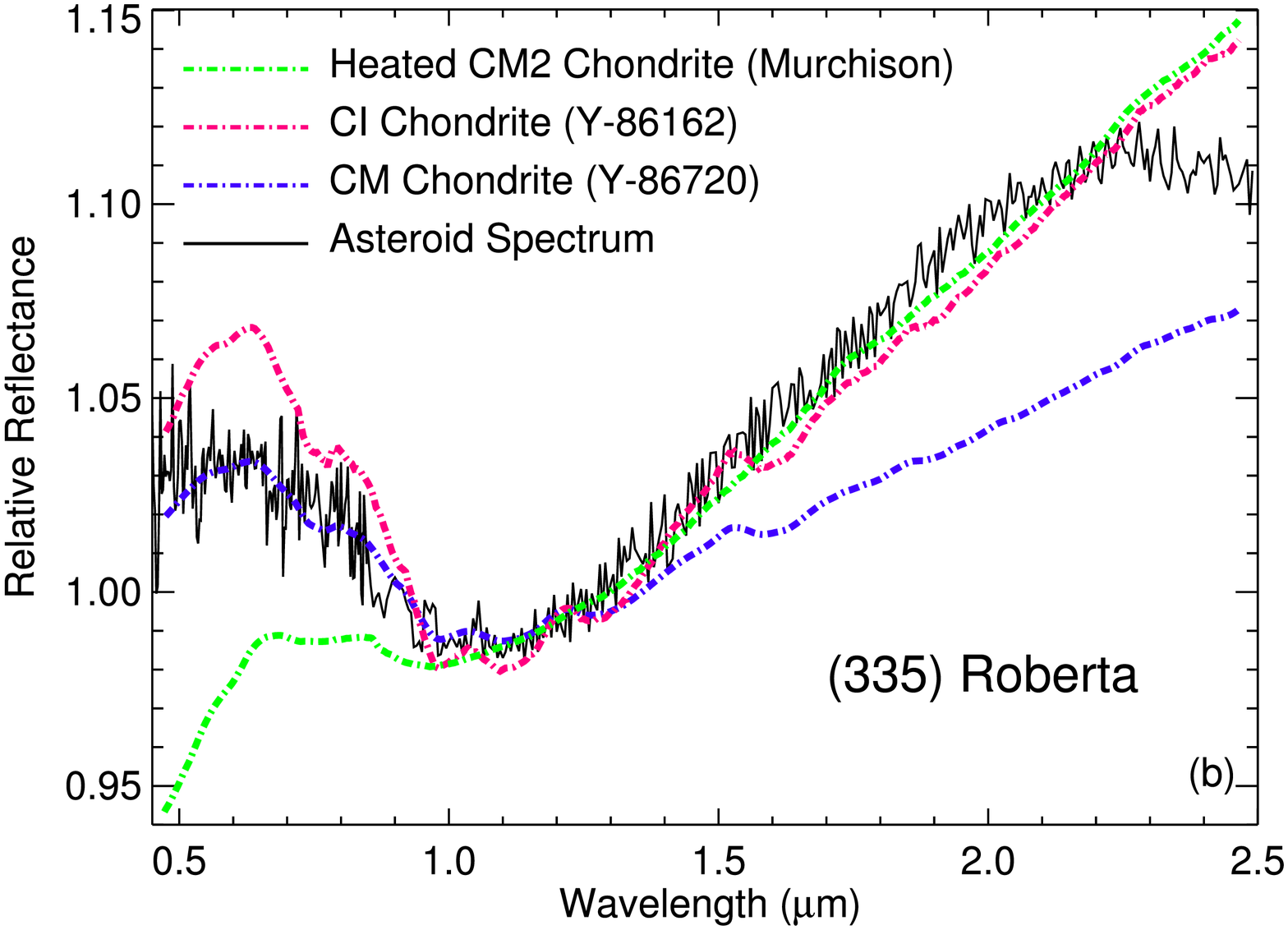}
\caption{Linear spectral mixing models, using laboratory spectra of silicates and carbonaceous chondrites, fit to the spectrum of B-type asteroid (335) Roberta. The Hedenbergite sample is from the USGS and the other samples are taken from RELAB. Panel (a) shows that the observed absorption feature can not be reproduced using silicates. The silicate spectra are scaled to fit the 1-$\mu$m band. Panel (b) shows that the CM chondrite model fits the data well at wavelengths shortwards of 1.3 $\mu$m but fails to match the asteroid spectrum at longer wavelengths. On the contrary, both the CI and the heated CM2 chondrite models fit the data better at longer wavelengths $\lambda > 1.0\ \mu$m, however significant discrepancies are observed below 1.0$\mu$m. }
\label{f2}
\end{center}
\end{figure}

We compared our data with laboratory spectra from two digital spectral libraries, namely the Brown University Keck/NASA Relab Spectra Catalog and the USGS Digital Spectral Library \citep{clark:2003}. Among well-studied laboratory samples, silicates (i.e.\ olivine and pyroxene) are well known for showing an absorption band centered near 1.0 $\mu$m. Areal mixing spectral models incorporating silicates and spectrally neutral materials (such as amorphous carbon), compared to the asteroid spectra, are shown in Figure \ref{f2}a.  The pyroxene model (shown as a blue dashed line in Figure \ref{f2}a) failed to fit the asteroid spectrum because it produces absorption bands at 0.9 $\mu$m and 1.8 $\mu$m neither of which is observed in the asteroid spectrum. The olivine model, on the other hand, matches the asteroid absorption feature moderately well from 0.7 $\mu$m to 1.4 $\mu$m. However, there is a discrepancy between the olivine model and the data at wavelengths $\lambda > $ 1.5 $\mu$m, becoming worse as wavelength increases. In addition, type A clinopyroxene \citep{adams:1975} was examined. This silicate sample exhibits a rather different 1$\mu$m absorption band, in terms of the spectral profile, compared with those of olivine and pyroxene. We searched for all available spectra of clinopyroxenes in the RELAB and the USGS libraries and fit the laboratory spectra to the asteroid spectra. Our best-fit clinopyroxene model is shown as the orange dashed line in Figure \ref{f2}a. We note that the 1-$\mu$m band of clinopyroxenes consists of two substructures \citep{Schade:2004}, which are not observed in the asteroid spectrum. As such, none of the silicate models is considered as a plausible spectral analog to the B-type asteroids. 

Alternatively, we examined carbonaceous chondrite samples. It has long been noticed that reflectance spectra of carbonaceous chondrites are similar to those of the low-albedo asteroids \citep{gaffey:1978, bell:1989, gaffey:1993, hiroi:1993, hiroi:1996}. We searched for spectral analogs of our B-type asteroids among available carbonaceous chondrite samples in the RELAB database, finding that CI and CM chondrites can better fit the asteroid spectra than other chondrite samples. Also, previous work by \cite{hiroi:1996} has shown that spectra of heated CM2 (Murchison) chondrite are similar to the spectra of some B-type asteroids. The  results using CI/CM and heated Murchison samples are illustrated in Figure \ref{f2}b. We found that even the best match among CM chondrites, namely Yamada-86720 (shown as the blue dashed line), provides only a marginal fit to the data in the optical region (0.5 - 0.9 $\mu$m) and it fits the asteroid spectrum rather poorly in the NIR. In contrast to CM chondrites, one unusual CI chondrite (Yamada-82162), shown as the pink dashed line in Figure \ref{f2}b, turned out to be a much better spectral analog of the B-type asteroids. Not only does the overall spectral profile of the Yamada-82162 chondrite generally match the shape of the asteroid spectrum, but also it fits the 1-$\mu$m absorption feature both in the band width and the band center. However, a noticeable difference between the asteroid spectrum and that of the CI chondrite appears in the visible wavelength regime, where the reflectivity of the meteorite rises up steeply and then shows a strong UV drop-off ($\lambda < 0.65 \mu$m) that is not present in the asteroid spectrum. Similarly, the heated Murchison model, shown in green, can only fit the asteroid spectrum at longer wavelengths and failed at wavelengths shortwards of 1.0 $\mu$m. 

\subsubsection{Detection of Magnetite}
\begin{figure}[h]
\begin{center}

\includegraphics[width=7in]{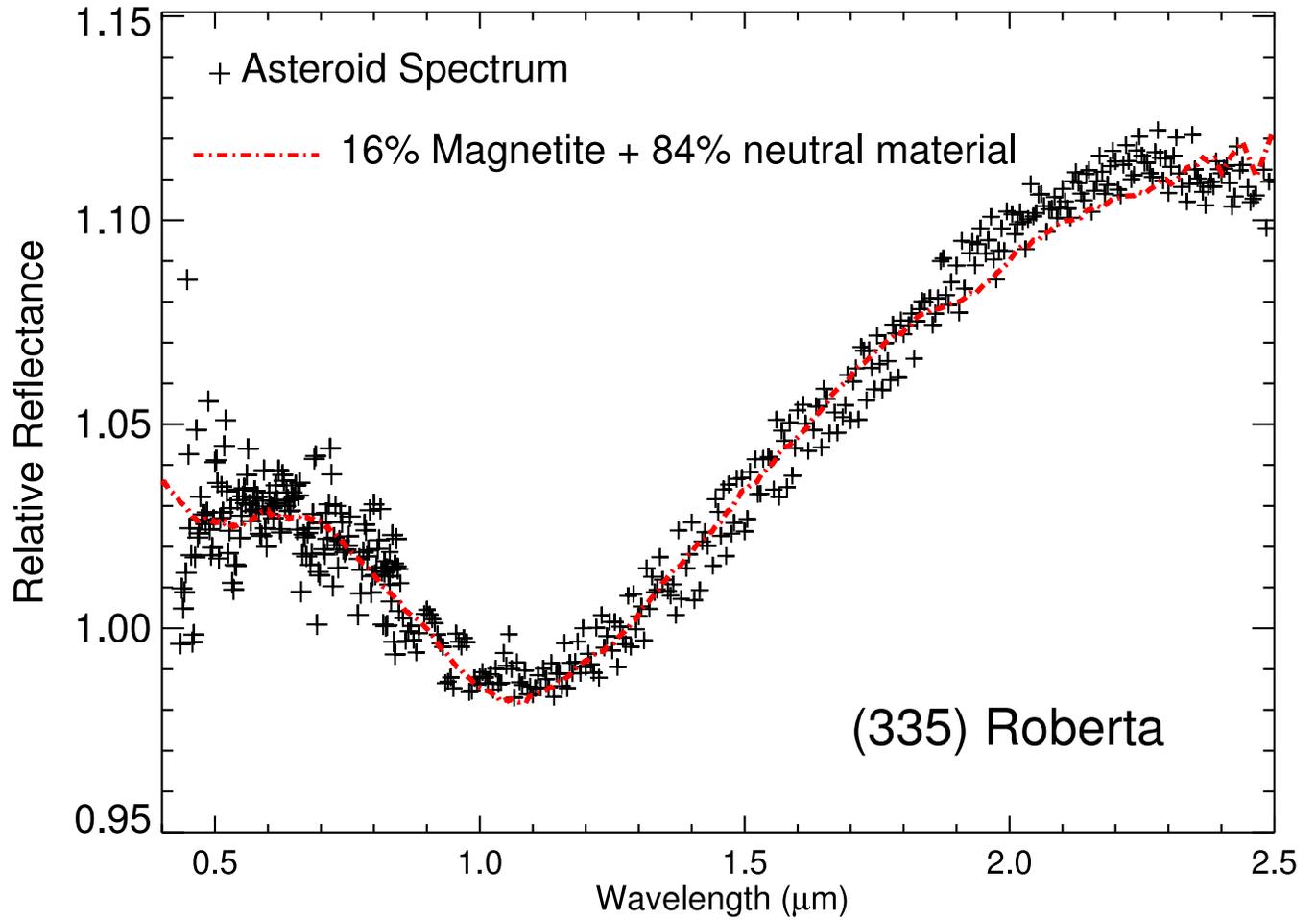}
\caption{Reflection spectrum of asteroid (335) Roberta compared with an areal mixing  model using 16\%, magnetite and 84\% spectrally neutral material.}
\label{f3}
\end{center}

\end{figure}

Surprisingly, we found that a mixture of magnetite (Fe$_{3}$O$_{4}$) (grain size 74 $\mu$m $< D <$ 250 $\mu$m, taken from the USGS spectral library) and spectrally neutral materials provides the best spectral match to several B-type asteroids. In particular, our areal mixing model, with 16\% magnetite and 84\% neutral material by area, fits the spectrum of asteroid (335) Roberta almost perfectly, see Figure \ref{f3}. The close match between magnetite and the asteroid Roberta suggests that magnetite may be a significant component of the surface materials. This dark iron oxide is abundant in carbonaceous chondrites and has been studied by several meteorite researchers \citep{Kerridge:1979,hyman:1983}. In chondritic meteorites, the presence of magnetite is an important indicator of the past aqueous alteration of its host body \citep{zolensky:1997}. If (335) Roberta had undergone aqueous alteration, then it should contain other mineralogical products, such as phyllosilicates (hydrated silicates).

Hydrous minerals have been reported to give rise to several absorption features observed in asteroids in the visible and the NIR. For example, a weak absorption band centered near 0.7 $\mu$m is attributed to an Fe$^{2+}$ $\rightarrow$ Fe$^{3+}$ charge transfer transition in hydrated minerals \citep{vilas:1989,vilas:1994}. In the NIR, overtone modes of bonded H$_{2}$O/OH in hydrous minerals produce two sharp absorption features at 1.4 and/or 1.9 $\mu$m, respectively \citep{bishop:1995}. We searched for these hydration features in the spectrum of Roberta. Even with the moderately high signal-to-noise ratio (SNR) data, we found neither the 0.7-$\mu$m band nor the 1.4 and 1.9 $\mu$m absorption features in this object. However, the intensity of these overtone bands and of the charge transfer feature, in the visible and NIR, is generally weak (at the level of a few percent), making these features susceptible to masking by opaque materials. Partly for this reason, previous studies have suggested that strong constraints on the hydration status of an asteroid are best obtained from observations at longer wavelengths, in particular near 3$\mu$m where the O-H bond has its fundamental transition \citep{rivkin:2002}.  

\section{3-$\mu$m Spectra}
The 3 $\mu$m region is especially important in the search for hydration features because a combination of bending and stretching modes of the O-H bond in water or hydroxyl-bearing minerals produces several diagnostic bands in this region. For instance, a symmetric stretch mode of the water molecule in ice results in a strong absorption feature at 3.1 $\mu$m and an asymmetric feature of O-H occurs at 2.9 $\mu$m \citep{aines:1984}. Unlike the absorption features at shorter wavelengths, the 3 $\mu$m band is commonly saturated and is detectable even with small ($\sim$2 wt\%) concentrations of water \citep{jones:1990}. As such, the fundamental absorption bands in the 3-$\mu$m region are difficult to suppress by absorbing constituents, and the 3-$\mu$m region can set strong constraints on the presence of hydrated minerals the target object. 

\subsection{Observation and Data Reduction}
We took  3-$\mu$m observations (1.9 - 4.2 $\mu$m) of asteroid (335) Roberta using the long wavelength cross-dispersed (LXD) Mode of SpeX on IRTF in May 2008. Additionally, the Infrared Camera and Spectrograph (IRCS) on the Subaru telescope \citep{kobayashi:2000} was used to observe (335) Roberta in the K- and L-bands in June 2008. For the Subaru observations, the low resolution Amici prism was adopted. The latter is designed specifically for 3-micron observations with high throughput and wide wavelength coverage (2.0 - 4.2 $\mu$m) for spatial scale 0.$\!\!^{\prime\prime}$020 pixel$^{-1}$ \citep{takato:2006}. The LXD data were reduced using the Spextool software with a method described in \citet{yang:2009}. The Subaru data were processed following the standard spectroscopic data reduction procedures based on the Image Reduction and Analysis Facility (IRAF), which are described in the Subaru data-reduction cookbook on the IRCS website\footnotemark. Spectra of the scientific targets and the standard stars were extracted from flattened, bad-pixel corrected and sky-subtracted images. Consequently, a set of individual spectra of the same object were combined to improve SNR. Since no good reference lamp is available for wavelength calibration in the L band, we used the transmission features of the atmosphere of the Earth in the standard star spectrum. Specifically, four telluric features in the spectrum of the standard star were used to define a second-order polynomial function to transform the pixel coordinate to the absolute wavelength. 

\footnotetext{http://subarutelescope.org/Observing/DataReduction/}

\subsection{Results} 
\begin{figure}[h]
\begin{center}
\vspace{1 cm}
\includegraphics[width=6in,angle=0]{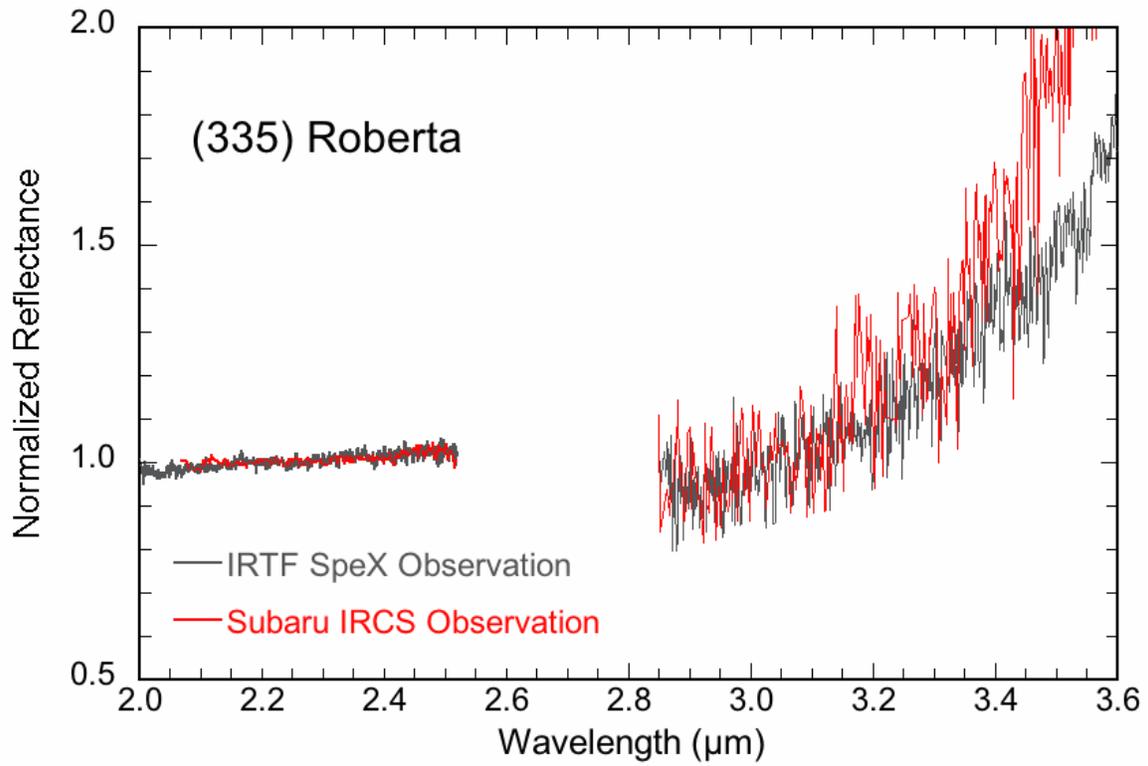}
\caption{K and L-band spectra of asteroid (335) Roberta. The IRTF observation (black line) was made on UT May 12, 2008. The Subaru observation (red line) was made on UT June 22, 2008. The slight divergence of the data sets at wavelengths $>$3.4 $\mu$m is due to the slightly higher effective temperature of Roberta when the Subaru data were taken. }
\label{f4}
\end{center}
\end{figure}

The SpeX spectrum (shown in black) and the IRCS spectrum (shown in red) of (335) Roberta are plotted in Figure \ref{f4}. The wavelength region from 2.5 to 2.8 $\mu$m is severely contaminated by telluric absorption, therefore, this region is omitted from Figure \ref{f4}. Although the two spectra were taken with different instruments on different telescopes, the LXD data and the IRCS data have almost the same wavelength coverage and show similar spectral profiles. In both spectra, a rapid increase in reflectivity beyond 3.1 $\mu$m was observed. At the time of the observations, asteroid Roberta was about 2 AU from the Sun (see Table \ref{obs}), suggesting that the steep increase in reflectivity is caused by thermal emission. We note that the Subaru spectrum has its thermal excess starting at a shorter wavelength than  in the IRTF spectrum. This may be because the Subaru data were taken when the asteroid was slightly closer to the Sun (2.14 AU vs. 2.21 AU, see Table \ref{obs}) and had higher surface temperature. 

We applied a simple thermal model to fit the thermal excess and removed the modeled thermal flux from the data. The thermal flux density is calculated by:

\begin{equation}
f_{BB}(\lambda) = \frac{\epsilon \pi B_\lambda(T_e) R_{e}^{2}}{\Delta^2}
\end{equation}

\noindent where $R_{e}$ is the effective radius of the object and $B_{\lambda}(T_c)$ is the Planck function evaluated at the effective temperature and $\epsilon$ is the emissivity. Given that no precise shape measurements are available for (335) Roberta, we assume that the asteroid is spherical with $R_{e}$  = 44.5 km \citep{tedesco:2002}. Previous studies have shown that the emissivity for rocks is close to unity at these wavelengths \citep{morrison:1973, fernandez:2003}, therefore we assume $\epsilon$ = 1.0. 

\begin{figure}[h]
\begin{center}
\vspace{1 cm}
\includegraphics[width=6.5in,angle=0]{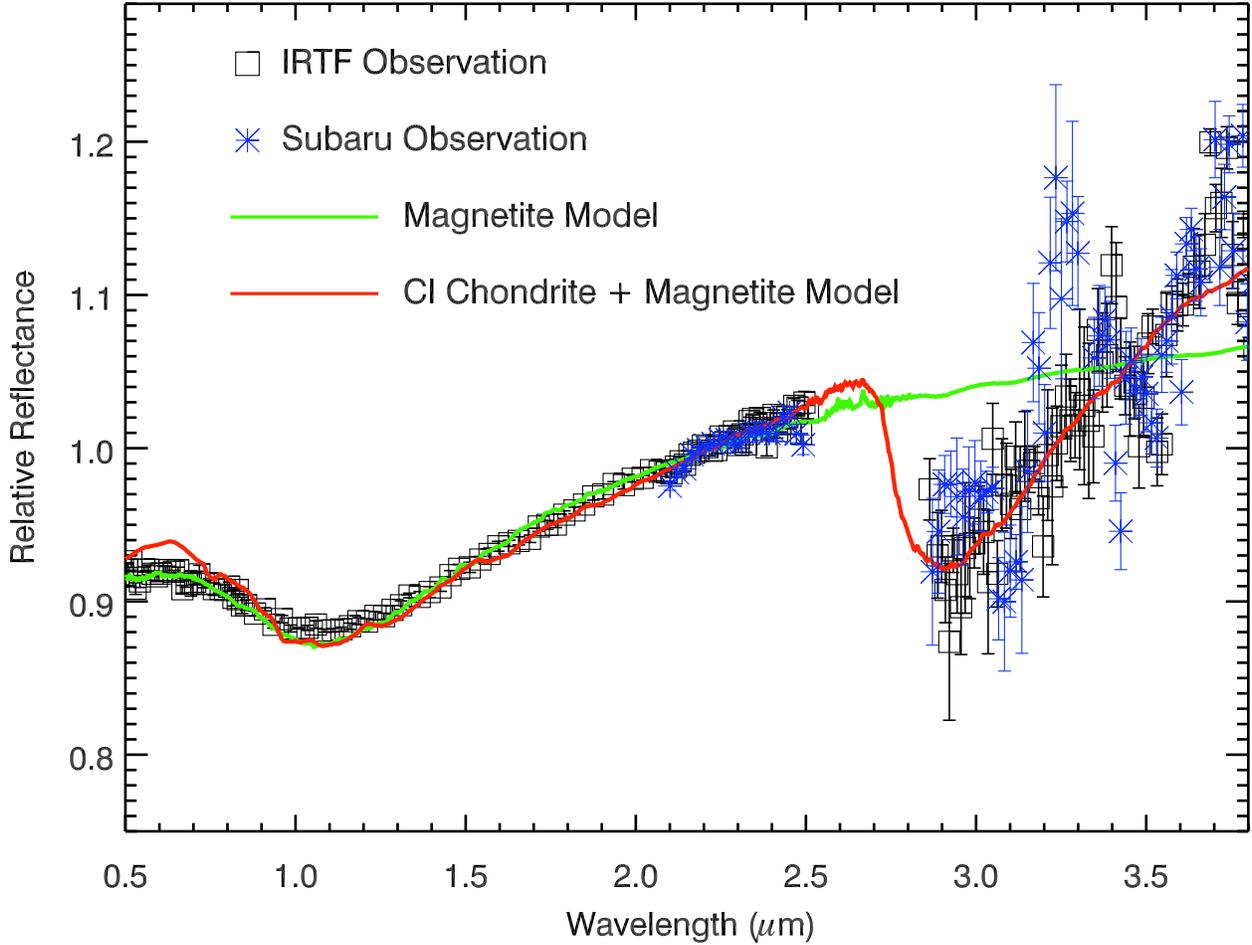}
\caption{The  IRTF spectrum of asteroid (335) Roberta is shown (black), from a combination of prism and LXD-mode observations with SpeX. The Subaru spectrum is also shown  (blue).  In both spectra, thermal emission has been subtracted as described in the text.  Both 1.0 and 2.9 $\mu$m absorption features were observed at the level of $\sim$ 10\% in both data sets.  The 1.0 $\mu$m band is consistent with magnetite (green line), but magnetite alone cannot fit the 2.9 $\mu$m band.  The full spectrum can be matched with a mixture of magnetite and  CI chondrite (red line).  
}
\label{f5}
\end{center}

\end{figure}

Our thermal model shows that the effective temperatures of Roberta were 235 $\pm$ 15K and 240 $\pm$ 20K at the time of the IRTF observation and the Subaru observation, respectively. The reflectance spectra of Roberta, after removing the thermal emission, are shown in Figure \ref{f5}, in which the IRTF data are shown as black open squares and the Subaru data are shown as blue stars. In order to enhance the SNR, we resampled the spectra in coarser wavelength bins of width 0.015 $\mu$m. An emission-like feature was observed between 3.3 and 3.4 $\mu$m in the Subaru data. This feature is caused by residuals from telluric absorption. Apart from this feature and other imperfect telluric subtractions, the IRTF data and the Subaru data are consistent with each other. An absorption feature at the level of $\sim$ 10\% was observed in both data sets. The band center and the spectral profile of the observed absorption feature are consistent with those of the diagnostic 2.9 $\mu$m band of phyllosilicates. Similar features have been observed in other low-albedo C-complex asteroids, which are rich in hydrated minerals \citep{rivkin:2003}. In particular, the shape of the 2.9-$\mu$m absorption feature (20\% in depth) in the spectrum of (2) Pallas \citep{rivkin:2003} is similar to that of Roberta. On the other hand, the depth of the absorption band can potentially be affected by telluric contamination, thermal removal or rotation. Therefore, additional observations over the entire rotational period of (335) Roberta are needed to better determine the strength of the 2.9 $\mu$m absorption band. 

Two linear areal-mixing spectral models are used to further investigate compositional constraints from the observations in the 3-$\mu$m region. The models and the data are presented in Figure \ref{f5}. The magnetite model, shown in green, fits the data in the visible and the NIR but it fails to produce the observed absorption band centered at 2.9 $\mu$m.  The band center is too short to be explained by water ice, as reported recently on (24) Themis (\cite{campins:2010},\cite{rivkin:2010}).  On the other hand, this 2.9 $\mu$m feature is consistent with the O-H absorption band commonly seen in laboratory spectra of carbonaceous chondrites (e.g. CM, CI). However, as we discussed earlier, the spectral models with CM/CI chondrites alone can not fit the data well at shorter wavelengths. As a compromise, we found that mixing CI chondrite with a small amount ($\sim$ 2\%) of pure magnetite improves the fit dramatically. The mixture spectrum, shown as solid red line, can fit the data successfully from the optical to the 3 $\mu$m region. 

\section{Discussion}
Our spectral models show that the synthetic spectrum of the mixture of Y-82162 chondrite and magnetite fits the spectrum of the asteroid (335) Roberta over a wide wavelength range. Given the high signal-to-noise ratio and spectral resolution of our data, we believe that the fit is significant, in the sense that the asteroid analogs and materials required to reproduced the spectrum are plausible constituents of the asteroids. Moreover, since only one model provides a high precision fit, the model strongly implies a specific mineralogy of the asteroid. 
 
Antarctic carbonaceous chondrite Y-82162 is classified into the CI group based on its oxygen isotopic properties \citep{kojima:1987} and magnetite is found to be an abundant phase in carbonaceous chondrites, up to 16\% by weight \citep{hyman:1982}.  \cite{Kerridge:1979} pointed out that the composition and morphology of magnetite in CI chondrites is incompatible with a nebular origin. The authors therefore suggested that magnetite was formed during a secondary mineralization on a planetesimal after the primitive condensation epoch \citep{Kerridge:1979}. Converging lines of evidence show that liquid water was widely available during the earliest geological era of the solar system (\cite{brearley:2006}, \cite{jewitt:2007}). Chemical reactions between water and rocks at temperatures perhaps only slightly above freezing lead to aqueous alteration that largely affected the oxidation states, mineralogy and surface chemistry of the host objects. In chondritic meteorites, the mineralogical products of aqueous alteration include hydrous minerals (i.e. serpentines and clays), hydroxides, oxides (magnetite and ferrihydrite), and carbonates \citep{brearley:2006}. The detections of magnetite, and of the 2.9 $\mu$m band in the spectrum of (335) Roberta, are consistent with aqueous alteration. Therefore, our observations suggest that B-type asteroid Roberta was once rich in water ice and had experienced intensive aqueous alteration. 

Many meteorites are derived from near-earth-objects (NEOs) \citep{mcsween:2006}. However, NEOs are dynamically unstable and short lived, so the NEO population must be continually replenished from other sources, which are the parent bodies of these meteorites. At present, the parent body for carbonaceous chondrites is under debate. While the widely held belief is that carbonaceous chondrites originate from asteroidal objects, \citep{lodders:1999} argued that the mineralogy and chemical characteristics of CI and CM chondrites are consistent with a cometary origin. However, the big problem for the cometary-parent scenario lies in the fact that CI and CM chondrites both have experienced intensive aqueous alteration, given the presence of rich hydrous minerals, carbonates, oxides and sulfates. Based on the latest studies of particles from the comet 81P/Wild 2 from the Stardust mission, no phyllosilicates and carbonates were found in the Wild 2 samples \citep{zolensky:2008}.  Although \cite{lisse:2006} reported detections of phyllosilicates and carbonates in the ejecta of Deep Impact, other secondary minerals (such as hydroxides, hydrated sulfides and oxides) are notably absent from comet 9P/Tempel 1\citep{lisse:2007}, 1P/Halley  \citep{Jessberger:1999} and 81P/Wild 2 \citep{zolensky:2006,zolensky:2008}. In particular, no magnetite (an important product of parent-body aqueous alteration) has ever been detected in comets \citep{abell:2005, campins:2006, campins:2007}. In fact, theoretical simulations have shown that is it more difficult for comets to host liquid water in their interiors because comets are likely to have accreted more amorphous ice than asteroids. Heating from $^{26}$Al would first crystallize amorphous ice and the excess heat could quickly escape from the comet interior due to the higher heat conductivity of crystalline ice \citep{Prialnik:2008, kelley:2009}. 

In contrast to comet observations, our detections of magnetite features in the NIR and phyllosilicate absorption band in the 3-$\mu$m region based on observations of B-type asteroid (335) Roberta reinforce the link between the carbonaceous chondrites and the main belt asteroids. As such, our observations suggest that this object is rich in magnetite and phyllosilicates and this object may contain parental materials for the CI chondrite.   

Lastly, we were motivated to study B-type asteroids by the observation that several meteor-stream-associated asteroids are blue. However, we found no evidence for water ice in B-type asteroids and so no evidence that they could ever lose mass though sublimation. Even more surprisingly, the NIR spectrum of Geminid parent (3200) Phaethon \citep{licandro:2007} shows no absorption band near 1 $\mu$m, and so differs from the B-types presented here. As such, a reasonable conclusion is that B-type asteroids are a heterogeneous group (c.f.  \cite{clark:2009}).

\section{Summary}
Our spectroscopic study of four B-type asteroids, namely (47) Aglaja, (142) Polana, (335) Roberta and (1615) Bardwell, in the infrared yields the following results:

\begin{itemize}
  \item A broad absorption feature, centered near 1.0 $\mu$m, was observed in the reflectance spectra of these asteroids. The absorption feature is consistent in position and shape with the presence of magnetite. 
  
  \item The  short-wavelength wing of the 1 $\mu$m band  is responsible for the principal defining characteristic - the blue optical color - of these B-type asteroids.   
  
   \item B-type asteroid (335) Roberta, which shows the 1 $\mu$m magnetite feature, also shows a band $\sim$10\% deep centered near 2.9 $\mu$m. We attribute this band to O-H bond vibrations in a hydrated mineral. The good magnetite-fit and the detection of 2.9 $\mu$m absorption strongly suggest that (335) Roberta  experienced significant aqueous alteration. 
\end{itemize}

\section{Acknowledgment}
We thank Alan Tokunaga for his support in granting the IRTF telescope time for this work and Drs. Sasha Krot,  Gary Huss and Ed Scott for valuable discussions and constructive suggestions. The authors also would like to acknowledge the RELAB assistance of Takahiro Hiroi. BY was supported by the National Aeronautics and Space Administration through the NASA Astrobiology Institute under Cooperative Agreement No. NNA08DA77A issued through the Office of Space Science and by a grant to David Jewitt from the NASA Planetary Origins program.


\begin{thebibliography}{47}
\expandafter\ifx\csname natexlab\endcsname\relax\def\natexlab#1{#1}\fi
\bibitem[{Abell {et~al.}(2005)Abell, Fern{\'a}ndez, Pravec, French, Farnham,
  Gaffey, Hardersen, Ku{\v s}nir{\'a}k, {\v S}arounov{\'a}, Sheppard, \&
  Narayan}]{abell:2005}
Abell, P.~A., {et~al.} 2005, Icarus, 179, 174

\bibitem[{Adams (1975)}]{adams:1975}
Adams, 1975. J.B. Adams, Interpretation of visible and near-infrared diffuse reflectance spectra of pyroxenes and other rock-forming minerals. In: C. Karr, Editor, Infrared and Raman Spectroscopy of Lunar and Terrestrial Minerals, Academic Press, New York (1975), pp. 91Ð116.

\bibitem[{Aines \& Rossman(1984)}]{aines:1984}
Aines, R.~D., \& Rossman, G.~R. 1984, Journal of Geophysical Research, 89, 4059

\bibitem[{Bell(1989)}]{bell:1989}
Bell, J.~F. 1989, Icarus, 78, 426

\bibitem[Binzel et al.(2002)]{binzel:2002} Binzel, R. P., Lupishko, D., di Martino, M., Whiteley, R. J., \& Hahn, G. J. 2002, in Asteroids III, ed. W. F. Bottke, Jr., et al. (Tucson: Univ. Arizona Press), 255

\bibitem[{Bishop \& Pieters(1995)}]{bishop:1995}
Bishop, J.~L., \& Pieters, C.~M. 1995, Journal of Geophysical Research, 100,
  5369

\bibitem[{Brearley(2006)}]{brearley:2006}
Brearley, A.~J. 2006, The Action of Water. In ``Meteorites and the Early Solar System II" , ed. Dante Lauretta, H.Y. McSween Jr and L. Leshin, (Arizona University Press), 587

\bibitem[Browning(1997)]{browning:1997} Browning, L., 1997, in Proceedings of Workshop on Parent-Body and Nebular Modification of Chondritic Materials, M.E. Zolensky, A.N. Krot and E.R.D. Scott (eds.), Lunar and Planetary Institute, 9

\bibitem[{Bus(1999)}]{bus:1999}
Bus, S.~J. 1999, Ph.D. thesis, MIT

\bibitem[{Bus \& Binzel(2002)}]{bus:2002}
Bus, S.~J., \& Binzel, R.~P. 2002, Icarus, 158, 146


\bibitem[{Campins {et~al.}(2006)Campins, Ziffer, Licandro, Pinilla-Alonso,
  Fern{\'a}ndez, Le{\'o}n, Moth{\'e}-Diniz, \& Binzel}]{campins:2006}
Campins, H., Ziffer, J., Licandro, J., Pinilla-Alonso, N., Fern{\'a}ndez, Y.,
  Le{\'o}n, J.~d., Moth{\'e}-Diniz, T., \& Binzel, R.~P. 2006, \aj, 132, 1346

\bibitem[{Campins {et~al.}(2007)Campins, Licandro, Pinilla-Alonso, Ziffer,
  de~Le{\'o}n, Moth{\'e}-Diniz, Guerra, \& Hergenrother}]{campins:2007}
Campins, H., Licandro, J., Pinilla-Alonso, N., Ziffer, J., de~Le{\'o}n, J.,
  Moth{\'e}-Diniz, T., Guerra, J.~C., \& Hergenrother, C. 2007, \aj, 134, 1626

\bibitem[Campins et al.(2010)]{campins:2010} Campins, H., et al.\ 
2010, \nat, 464, 1320 

\bibitem[{Cellino {et~al.}(2001)Cellino, Zappal{\`a}, Doressoundiram,
  di~Martino, Bendjoya, Dotto, \& Migliorini}]{Cellino:2001}
Cellino, A., Zappal{\`a}, V., Doressoundiram, A., di~Martino, M., Bendjoya, P.,
  Dotto, E., \& Migliorini, F. 2001, Icarus, 152, 225

\bibitem[{Chamberlin {et~al.}(1996)Chamberlin, McFadden, Schulz, Schleicher, \&
  Bus}]{chamberlin:1996}
Chamberlin, A.~B., McFadden, L.-A., Schulz, R., Schleicher, D.~G., \& Bus,
  S.~J. 1996, Icarus, 119, 173


\bibitem[{Clark {et~al.}(2003)Clark, Swayze, Wise, Livo, Hoefen, Kokaly, \&
  Sutley}]{clark:2003}
Clark, R.~N., Swayze, G.~A., Wise, R., Livo, K.~E., Hoefen, T.~M., Kokaly,
  R.~F., \& Sutley, S.~J. 2003, in American Astronomical Society, 35th DPS
  meeting, Vol.~35
  
  \bibitem[Clark et al.(2009)]{clark:2009} Clark, B.~E., et al.\ 
2009, AAS/Division for Planetary Sciences Meeting Abstracts, 41, \#32.08 


\bibitem[Cunningham(1950)]{cunningham:1950} Cunningham, L.~E.\ 1950, 
\iaucirc, 1250, 3

\bibitem[{Cushing {et~al.}(2004)Cushing, Vacca, \& Rayner}]{cushing:2004}
Cushing, M.~C., Vacca, W.~D., \& Rayner, J.~T. 2004, Publications of the
  Astronomical Society of the Pacific, 116, 362

\bibitem[Dandy et al.(2003)]{dandy:2003} Dandy, C.~L., 
Fitzsimmons, A., \& Collander-Brown, S.~J.\ 2003, Icarus, 163, 363 

\bibitem[Fernandez et al.(1997)]{fernandez:1997} Fern\`andez, Y.~R., 
McFadden, L.~A., Lisse, C.~M., Helin, E.~F., 
\& Chamberlin, A.~B.\ 1997, Icarus, 128, 114 

\bibitem[{Fern\`andez {et~al.}(2003)Fern\`andez, Sheppard, \&
  Jewitt}]{fernandez:2003} Fern\`andez, Y.~R., Sheppard, S.~S., \& Jewitt, D.~C. 2003, \aj, 126, 1563

\bibitem[{Gaffey {et~al.}(1993)Gaffey, Burbine, \& Binzel}]{gaffey:1993}
Gaffey, M.~J., Burbine, T.~H., \& Binzel, R.~P. 1993, Meteoritics, 28, 161

\bibitem[{Gaffey \& McCord(1978)}]{gaffey:1978}
Gaffey, M.~J., \& McCord, T.~B. 1978, Space Science Reviews, 21, 555

\bibitem[{Green \& Kowal(1983)}]{green:1983}
Green, S., \& Kowal, C. 1983, IAU Circ., 3878

\bibitem[{Hiroi {et~al.}(1993)Hiroi, Pieters, Zolensky, \&
  Lipschutz}]{hiroi:1993} Hiroi, T., Pieters, C.~M., Zolensky, M.~E., \& Lipschutz, M.~E. 1993, Science,
  261, 1016

\bibitem[{Hiroi {et~al.}(1996)Hiroi, Zolensky, Pieters, \&
  Lipschutz}]{hiroi:1996}
Hiroi, T., Zolensky, M.~E., Pieters, C.~M., \& Lipschutz, M.~E. 1996,
  Meteoritics and Planetary Science, 31, 321
  
\bibitem[Hsieh 
\& Jewitt(2005)]{hsieh:2005} Hsieh, H.~H., \& Jewitt, D.\ 2005, \apj, 624, 1093 

\bibitem[{Hsieh \& Jewitt(2006)}]{hsieh:2006}
Hsieh, H.~H., \& Jewitt, D. 2006, Science, 312, 561

\bibitem[{Hyman \& Rowe(1982)}]{hyman:1982}
Hyman, M., \& Rowe, M.~W. 1982, in Lunar and Planetary Science XIII, 354

\bibitem[{Hyman \& Rowe(1983)}]{hyman:1983}
Hyman, M., \& Rowe, M.~W. 1983, Journal of Geophysical Research, 88

\bibitem[{Jenniskens {et~al.}(1998)Jenniskens, Betlem, de~Lignie, Ter~Kuile,
  van Vliet, van~`t Leven, Koop, Morales, \& Rice}]{Jenniskens:1998}
Jenniskens, P., {et~al.} 1998, Monthly Notices of the Royal Astronomical
  Society, 301, 941
  
  \bibitem[Jenniskens(2008)]{2008EM&P..102..505J} Jenniskens, P.\ 2008, Earth Moon and Planets, 102, 505 

\bibitem[{Jessberger(1999)}]{Jessberger:1999}
Jessberger, E.~K. 1999, Space Science Reviews, 90, 91

\bibitem[Jewitt(2002)]{2002AJ....123.1039J} Jewitt, D.~C.\ 2002, \aj, 123, 
1039 

\bibitem[{Jewitt \& Hsieh(2006)}]{jewitt:2006}
Jewitt, D., \& Hsieh, H. 2006, Astronomical Journal, 132, 1624

\bibitem[Jewitt et al.(2007)]{jewitt:2007} Jewitt, D., Chizmadia, 
L., Grimm, R., \& Prialnik, D.\ 2007, Protostars and Planets V, University of Arizona Press, (ed: B. Reipurth et al.), 863 


\bibitem[{Jones {et~al.}(1990)Jones, Lebofsky, Lewis, \& Marley}]{jones:1990}
Jones, T.~D., Lebofsky, L.~A., Lewis, J.~S., \& Marley, M.~S. 1990, Icarus, 88,
  172

\bibitem[Kasuga 
\& Jewitt(2008)]{kasuga:2008} Kasuga, T., \& Jewitt, D.\ 2008, \aj, 136, 881

\bibitem[{Kelley \& Wooden(2009)}]{kelley:2009}
Kelley, M.~S., \& Wooden, D.~H. 2009, Planetary and Space Science, 57, 1133

\bibitem[{Kerridge {et~al.}(1979)Kerridge, Mackay, \& Boynton}]{Kerridge:1979}
Kerridge, J.~F., Mackay, A.~L., \& Boynton, W.~V. 1979, Science, 205, 395

\bibitem[{Kinoshita {et~al.}(2007)Kinoshita, Ohtsuka, Sekiguchi, Watanabe, Ito,
  Arakida, Kasuga, Miyasaka, Nakamura, \& Lin}]{kinoshita:2007}
Kinoshita, D., {et~al.} 2007, Astronomy and Astrophysics, 466, 1153

\bibitem[{Kobayashi {et~al.}(2000)Kobayashi, Tokunaga, Terada, Goto, Weber,
  Potter, Onaka, Ching, Young, Fletcher, Neil, Robertson, Cook, Imanishi, \&
  Warren}]{kobayashi:2000}
Kobayashi, N., {et~al.} 2000, in Optical and IR Telescope Instrumentation and Detectors, M. Iye, A. F. Moorwood (Eds). Proc. SPIE Vol. 4008,1056

\bibitem[{Kojima \& Yanai(1987)}]{kojima:1987}
Kojima, H., \& Yanai, K. 1987, Meteoritics, 22

\bibitem[{Lazzaro {et~al.}(2004)Lazzaro, Angeli, Carvano, Motha-Diniz, Duffard,
  \& Florczak}]{lazzaro:2004}
Lazzaro, D., et al., 2004, Icarus, 172, 179

\bibitem[Lazzarin et al.(1996)]{lazzarin:1996} Lazzarin, M., Barucci, 
M.~A., \& Doressoundiram, A.\ 1996, Icarus, 122, 122 


\bibitem[{Licandro {et~al.}(2007)Licandro, Campins, Moth{\'e}-Diniz,
  Pinilla-Alonso, \& de~Le{\'o}n}]{licandro:2007}
Licandro, J., et al. 2007, Astronomy and Astrophysics, 461, 751


\bibitem[Licandro \& Campins (2009)]{licandro:2009} Licandro, J. \& Campins, H., (2009). Are the main belt comets, comets? Proceedings of the International Astronomical Union, 5, 215


\bibitem[{Lisse {et~al.}(2006)Lisse, VanCleve, Adams, A'Hearn, Fern{\'a}ndez,
  Farnham, Armus, Grillmair, Ingalls, Belton, Groussin, McFadden, Meech,
  Schultz, Clark, Feaga, \& Sunshine}]{lisse:2006} Lisse, C.~M., et al.\ 2006, Science, 313, 635

\bibitem[{Lisse {et~al.}(2007)}]{lisse:2007}Lisse, C.~M., Kraemer, 
K.~E., Nuth, J.~A., Li, A., \& Joswiak, D.\ 2007, Icarus, 187, 69 

\bibitem[{Lodders \& Osborne(1999)}]{lodders:1999}
Lodders, K., \& Osborne, R. 1999, Space Science Reviews, 90, 289

\bibitem[{Luu \& Jewitt(1990)}]{Luu:1990}
Luu, J.~X., \& Jewitt, D.~C. 1990, Astronomical Journal, 99, 1985

\bibitem[{Mason(1995)}]{Mason:1995}
Mason, J.~W. 1995, Journal of the British Astronomical Association, 105, 219

\bibitem[McKinley(1961)]{mckinley:1961}McKinley D. W. R., 1961, Meteor Science and Engineering. McGrawÐHill, New York

\bibitem[McSween et al.(2006)]{mcsween:2006} McSween, H.~Y., Jr., 
Lauretta, D.~S., \& Leshin, L.~A.\ 2006, Meteorites and the Early Solar System II, D. S. Lauretta and H. Y. McSween Jr. (eds.), University of Arizona Press, Tucson, 53

\bibitem[{Meng {et~al.}(2004)Meng, Zhu, Gong, Li, Yang, Gao, Guan, Fan, \&
  Xia}]{meng:2004}
Meng, H., {et~al.} 2004, Icarus, 169, 385

\bibitem[{Morrison(1973)}]{morrison:1973}
Morrison, D. 1973, Icarus, 19, 1

\bibitem[{Ohtsuka {et~al.}(2006)Ohtsuka, Sekiguchi, Kinoshita, Watanabe, Ito,
  Arakida, \& Kasuga}]{ohtsuka:2006}
Ohtsuka, K., {et~al.}, 2006, \aap, 450, L25

\bibitem[Ohtsuka et 
al.(2008)]{ohtsuka:2008} Ohtsuka, K., Arakida, H., Ito, T., Yoshikawa, M., \& Asher, D.~J.\ 2008, Meteoritics and Planetary Science Supplement, 43, 5055 

\bibitem[{Prialnik {et~al.}(2008)Prialnik, Sarid, Rosenberg, \&
  Merk}]{Prialnik:2008}
Prialnik, D., Sarid, G., Rosenberg, E.~D., \& Merk, R. 2008, Space Science
  Reviews, 138, 147

\bibitem[{Rayner {et~al.}(2003)Rayner, Toomey, Onaka, Denault, Stahlberger,
  Vacca, Cushing, \& Wang}]{rayner:2003}
Rayner, J.~T., Toomey, D.~W., Onaka, P.~M., Denault, A.~J., Stahlberger, W.~E.,
  Vacca, W.~D., Cushing, M.~C., \& Wang, S. 2003, Publications of the
  Astronomical Society of the Pacific, 115, 362

\bibitem[{Rivkin {et~al.}(2002)Rivkin, Howell, Vilas, \&
  Lebofsky}]{rivkin:2002}
Rivkin, A.~S., Howell, E.~S., Vilas, F., \& Lebofsky, L.~A. 2002, Asteroids
  III, W. F. Bottke Jr., A. Cellino, P. Paolicchi, and R. P. Binzel (eds), University of Arizona Press, Tucson, 235
  \bibitem[{Rivkin {et~al.}(2003)Rivkin, Davies, Johnson, Ellison, Trilling,
  Brown, \& Lebofsky}]{rivkin:2003}
Rivkin, A.~S., Davies, J.~K., Johnson, J.~R., Ellison, S.~L., Trilling, D.~E.,
  Brown, R.~H., \& Lebofsky, L.~A. 2003, Meteoritics and Planetary Science, 38,
  1383
  
  \bibitem[Rivkin 
\& Emery(2010)]{rivkin:2010} Rivkin, A.~S., \& Emery, J.~P.\ 2010, \nat, 464, 1322 



\bibitem[{Schade {et~al.}(2004)Schade, W{\"a}sch, \& Moroz}]{Schade:2004}
Schade, U., W{\"a}sch, R., \& Moroz, L. 2004, Icarus, 168, 80

\bibitem[Takato 
\& Terada(2006)]{takato:2006} Takato, N., \& Terada, H.\ 2006, \procspie, 6269.

\bibitem[{Tedesco {et~al.}(2002)Tedesco, Noah, Noah, \& Price}]{tedesco:2002}
Tedesco, E.~F., Noah, P.~V., Noah, M., \& Price, S.~D. 2002, \aj, 123, 1056

\bibitem[{Tholen(1984)}]{tholen:1984}
Tholen, D.~J. 1984, Ph.D. Thesis, Univ. of Arizona

\bibitem[{Vilas \& Gaffey(1989)}]{vilas:1989}
Vilas, F., \& Gaffey, M.~J. 1989, Science, 246, 790

\bibitem[{Vilas(1994)}]{vilas:1994}
Vilas, F. 1994, Icarus, 111, 456

\bibitem[Whipple(1951)]{whipple:1951} Whipple, F.~L.\ 1951, \apj, 
113, 464 

\bibitem[{Whipple(1983)}]{whipple:1983}
Whipple, F.~L. 1983, IAU Circ., 3881


\bibitem[{Yang {et~al.}(2003)Yang, Zhu, Gao, Ma, Zhou, Wu, \& Guan}]{yang:2003}
Yang, B., Zhu, J., Gao, J., Ma, J., Zhou, X., Wu, H., \& Guan, M. 2003,
  \aj, 126, 1086

\bibitem[{Yang {et~al.}(2009)Yang, Jewitt, \& Bus}]{yang:2009}
Yang, B., Jewitt, D., \& Bus, S.~J. 2009, \aj, 137, 4538

\bibitem[{Zappala {et~al.}(1995)Zappala, Bendjoya, Cellino, Farinella, \&
  Froeschle}]{Zappala:1995}
Zappala, V., Bendjoya, P., Cellino, A., Farinella, P., \& Froeschle, C. 1995,
  Icarus, 116, 291

\bibitem[Zolensky(1997)]{zolensky:1997} Zolensky, M.~E., 1997, in Proceedings of Workshop on Parent-Body and Nebular Modification of Chondritic Materials, M.E. Zolensky, A.N. Krot and E.R.D. Scott (eds.), Lunar and Planetary Institute, 9

\bibitem[{Zolensky {et~al.}(2006)}]{zolensky:2006}
Zolensky, M.~E., {et~al.} 2006, Science, 314, 1735

\bibitem[{Zolensky {et~al.}(2008)}]{zolensky:2008}
Zolensky, M. ~E., {et~al.} 2008, Meteoritics and Planetary Science, 43, 261


\end{thebibliography}

\begin{deluxetable}{lllrccccl}\tablewidth{5.5in}
\tabletypesize{\scriptsize}
\tablecaption{ Observational Parameters for the B-type Asteroids
  \label{obstable}}
\tablecolumns{6} \tablehead{ \colhead{Object}   &
\colhead{ Date } & \colhead{Obs.}& \colhead{Exp.} & \colhead{Airmass} & \colhead{$r$ \tablenotemark{a}}& \colhead{$\Delta$ \tablenotemark{b}}&\colhead{$\phi$ \tablenotemark{c}}& \colhead{Standard}\\
\colhead{} &\colhead{UT} & \colhead{Mode}& \colhead{sec.} & \colhead{} & \colhead{AU} & \colhead{AU} &\colhead{$^o$}& \colhead{}}
\startdata
47 Aglaja & 2008 May 12 & SpeX (Prism) & 720 & 1.50 &2.576& 1.940& 20.2& HD193193 \\
142 Polana & 2008 July 25 & SpeX (Prism) & 720 & 1.08 &2.482&1.674 &17.3& HD223228 \\
335 Roberta & 2008 May 12 & SpeX (LXD) & 360&1.29&2.207&1.208 & 5.3& HD140990 \\
335 Roberta & 2008 May 13 & SpeX (Prism) & 360&1.13&2.205&1.208 & 5.6& HD100038 \\
335 Roberta & 2008 June 22 & IRCS (Prism)&600&1.17& 2.143 &1.366 & 22.0& HD131876 \\
1615 Bardwell & 2008 July 25 & SpeX (Prism) &1200 &1.28&2.949&2.014& 9.3& HD208620 \\
\enddata

\tablenotetext{a}{Heliocentric distances are taken from JPL/Horizon. }
\tablenotetext{b}{Geocentric distances are taken from JPL/Horizon. }
\tablenotetext{c}{Phase angles at the time of the observations, taken from JPL/Horizon.}
\label{obs}
\end{deluxetable}

\begin{deluxetable}{lcccccccccccc}\tablewidth{3.9in}
\tabletypesize{\scriptsize}
\tablecaption{Orbital and Physical Properties of the B-types
  \label{obstable}}
\tablecolumns{8} \tablehead{ \colhead{Object} &\colhead{ $q$ \tablenotemark{a}}  &  \colhead{ $a$ \tablenotemark{b} } &  \colhead{ $i$ \tablenotemark{c} }  &  \colhead{$e$ \tablenotemark{d} } & \colhead{$D$ \tablenotemark{e}} & \colhead{$p_v$ \tablenotemark{f}} & \colhead{Center \tablenotemark{g}} \\
\colhead{}&\colhead{(AU)}&\colhead{(AU)}&\colhead{$^{o}$}&\colhead{}&\colhead{(km)}&\colhead{}&\colhead{$\mu$m}}
\startdata
47 Aglaja  & 2.503 & 2.881 &4.99 &0.13 &127 & 0.080 & 1.13  \\
142 Polana  & 2.087 & 2.418 &2.24 &0.14 & 55 & 0.045  & 1.12  \\
335 Roberta  & 2.042 & 2.474 &5.09 & 0.17& 89 & 0.058 & 1.17 \\
1615 Bardwell & 2.565 & 3.121 & 1.69 &0.18 & 28 & 0.064 & 1.17  \\
\enddata
\tablenotetext{a}{Perihelion distance in (AU). From JPL/Horizon.}
\tablenotetext{b}{Orbital semi-major axis in (AU). From JPL/Horizon.}
\tablenotetext{c}{Orbital inclination in (degree). From JPL/Horizon.}
\tablenotetext{d}{Orbital eccentricity. From JPL/Horizon.}
\tablenotetext{e}{Mean diameter computed from absolute magnitude (H) and IRAS albedo.}
\tablenotetext{f}{Mean geometric albedo based on IRAS photometry
 \citep{tedesco:2002}.}
\tablenotetext{g}{Center of the broad absorption band.}
\label{orb_phy}
\end{deluxetable}

\end{document}